\documentclass[twoside,twocolumn,english,prd, nofootinbib, showkeys, showpacs, longbibliography]{revtex4-1}
\usepackage[T1]{fontenc}
\usepackage{babel}
\usepackage{amsmath}
\usepackage{amsthm}
\usepackage{graphicx}
\usepackage{bm}
\usepackage[unicode=true,pdfusetitle, bookmarks=true,bookmarksnumbered=false,bookmarksopen=false, breaklinks=false,pdfborder={0 0 1},backref=false,colorlinks=false]{hyperref}
\usepackage{microtype}

\DeclareMathOperator{\upd}{\mathrm{d\!}}

\begin{document}

\title{Solution of Supplee's submarine paradox through special and general relativity}
\author{R. S. Vieira}
\email{rsvieira@df.ufscar.br}
\affiliation{Universidade Federal de São Carlos, Caixa postal 676, CEP 13565-905, São Carlos-SP, Brasil}
\begin{abstract}
In 1989 Supplee described an apparent relativistic paradox on which a submarine seems to sink to observers at rest within the ocean, but it rather seems to float in the submarine proper frame. In this letter, we show that the paradox arises from a misuse of the Archimedes principle in the relativistic case. Considering first the special relativity, we show that any relativistic force field can be written in the Lorentz form, so that it can always be decomposed into a \emph{static} (electric-like) and a \emph{dynamic} (magnetic-like) part. These gravitomagnetic effects provide a relativistic formulation of Archimedes principle, from which the paradox is explained. Besides, if the curved spacetime on the vicinity of the Earth is taken into account, we show that the gravitational force exerted by Earth on a moving body must increase with the speed of the body. The submarine paradox is then analyzed again with this speed-dependent gravitational force.
\end{abstract}
\maketitle

\section{The paradox}

When a submarine is submerged underwater it can sink or float, depending
on whether its density is higher or lower than the density of the
water. Suppose we adjust the submarine density to that of the ocean
water, when both of them are at rest, so that the submarine remains
in equilibrium when submerged. What should happen, then, when the
submarine is put to move with a high velocity in the water? Disregarding
any hydrodynamic effects as drag, viscosity, turbulence etc. (which
we shall always assume hereafter), observers fixed to the ocean would
claim that the submarine sinks, since its density becomes higher than
the water density thanks to the Lorentz contraction. On the other
hand, observers within the submarine would claim instead that the
submarine should float, since it is the water that now becomes denser.
Of course, the submarine cannot float in a frame and sink in another,
so we get a contradictory situation.

This apparent paradox was described in 1989 by Supplee in \cite{Supplee1989},
although he had used a bullet instead of a submarine. Considering
some assumptions about the gravitational force among moving bodies,
Supplee gave two explanations for the problem, from which he concluded
that the bullet should sink in both frames. In the first explanation,
he avoided to use the theory of gravitation by considering a uniformly
upwards accelerated lake which, according to Einstein's equivalence
principle, behaves like a uniform gravitational field; he showed
that in this case the bullet acceleration is less than that of the
lake, so that the bullet relatively sinks. In the bullet proper frame,
however, the water indeed becomes denser but, since this frame is
no longer inertial, the isobaric surfaces of the lake will not be
flat anymore, which ultimately results in the bullet going far away
from the lake surface, \emph{i.e.}, in the bullet sinking again. In
the second explanation, Supplee considered a constant weak field in
the framework of general relativity, which led him to the same conclusion.

Fourteen years after Supplee's publication, Matsas had analyzed the
problem again, but this time using the full machinery of general relativity
\cite{Matsas2003}. Considering a background spacetime with a Rindler
chart and assuming reasonable conditions about the submarine rigidness,
Matsas analyzed the motion of a submarine which accelerates from
rest to a given velocity $\boldsymbol{v}$. He concluded that the
submarine shape gets deformed as it accelerates, with its length contracting
more and more, so that its density increases accordingly, which leads
the submarine to sink. Moreover, in the proper frame of the submarine
he showed that the observed gravitational field is somewhat different, which leads the submarine to sink as well.
Matsas also argued that
this problem can be important to some questions regarding the thermodynamic of black-holes, for instance, the self-consistency of Bekenstein formulation of the second law of thermodynamics, where the buoyancy force induced by Hawking's radiation plays a significant role $-$ see \cite{Matsas2003} and references therein.

Finally, Supplee's paradox was studied once more by Jonsson through
the analysis of the fictitious forces that appear in non-inertial
frames \cite{Jonsson2006}. Jonsson considered both a flat as well
as a spherical ocean. Such a flat ocean can be thought, with sufficient
accuracy, as an ordinary ocean in the Earth's surface, while the spherical
ocean can be regarded as that present perhaps in a very dense planet
or surrounding the core of a black hole. In the first case of a flat
ocean, Jonsson concluded that the submarine indeed sinks but, in the
case of the spherical ocean, he argued that the submarine could sink
or float depending on whether it moves respectively inside or outside
the so called \emph{photon sphere} $-$ the spherical surface on which
light can travel in closed orbits \cite{Taylor2000,Misner1973,Landau1980,Claudel2001}.
We remark, however, that Jonsson second analysis seems to contradict
his first one: in fact, since Earth is not dense enough in order to
have an external photon sphere, we would conclude from Jonsson's second
analysis that the submarine should float in a flat ocean instead of
sinking.

Although the above mentioned approaches are interesting by themselves,
we believe that it is not necessary to employ accelerated frames nor to use the full theory of general relativity in order to explain the submarine paradox.
In fact, first of all, we should remark that accelerated
motions can be contemplated with special relativity without the use of
non-inertial frames, so the introduction of non-inertial frames to
explain the submarine paradox is not necessary. Moreover, the behavior
of the submarine $-$ if it sinks or floats $-$ depends only on the balance
between the Archimedes (buoyancy) force and the gravitational (weight)
force acting on it. On the surface of Earth, the gravitational field
is relatively very small (in the sense that the spacetime curvature can
be neglected for any practical purpose), which enable us to interpret
the gravitational interaction as an ordinary force field in a flat
spacetime. This, of course, is only an approximation, since general
relativity show us that in an exact flat spacetime there is no gravity.
Nevertheless, the spacetime in the vicinity of Earth can be regarded,
with a very high accuracy, as consisting of a \emph{flat space} plus
a \emph{curved time}. We shall show that even in this case the gravitational
field can still be interpreted as a force field, although it must
become dependent on the speed of the bodies. The special theory of relativity
can also be employed with some care in this case and, thus, the submarine
paradox can be explained in both a flat as well as in a curved spacetime.

To explain the submarine paradox with our approach, however, it will
be necessary to impose that the gravitational force is covariant under the Lorentz transformations. This led us to a \emph{covariant theory
of gravitation} in a flat spacetime, as described in \cite{Vieira2016}.
This theory also holds in a flat space plus a curved time when the
speed dependence of the gravitational force is taken into account.
This covariance requirement implies that gravitomagnetic effects,
which play a key hole in our explanation of the Supplee's submarine
paradox, must be present whenever there is a relative motion between
two or more interacting bodies.

\section{Any covariant force field of special relativity can be written in
Lorentz form\label{Sec: Lorentz}}

The special theory of relativity tells us that the force is not a four-vector.
In fact, if $\boldsymbol{F}=\upd \boldsymbol{p}/\upd t$ is the force acting
in a given body, as measured from an inertial frame $R$, and $\boldsymbol{F}'=\upd \boldsymbol{p}'/\upd t'$
is the same force but measured by another inertial frame $R'$ (with
$R'$ moving with respect to $R$ with the velocity $\boldsymbol{v}=v\hat{\boldsymbol{x}}$,
all the axes being coincident at $t=t'=0$), then we get that \cite{Resnick1968,French1968}\begin{subequations} \label{F}
\begin{align}
F_{x}' & =F_{x}-v/c^{2}\left(u_{y}F_{y}+u_{z}F_{z}\right)/\left(1-u_{x}v/c^{2}\right),\\
F_{y}' & =F_{y}/\left[\gamma\left(1-u_{x}v/c^{2}\right)\right],\\
F_{z}' & =F_{z}/\left[\gamma\left(1-u_{x}v/c^{2}\right)\right],
\end{align}
\end{subequations}
where $\gamma=\sqrt{1-v^2/c^2}$ and $\bm{u}$ is the velocity of the body as measured by $R$. In this section, we shall show that although the force is not a four-vector,
it can be always written in a Lorentz form,
\begin{equation}
\boldsymbol{F}=\boldsymbol{G}+\boldsymbol{u}\times\boldsymbol{H},\label{Lorentz}
\end{equation}
with respect to any inertial frame. In (\ref{Lorentz}), $\boldsymbol{G}$
is defined as the part of the force $\boldsymbol{F}$ which does not
depend on the body velocity $\boldsymbol{u}$ $-$ we may call it
the \emph{static} (electric-like) part of the force. Similarly, $\boldsymbol{M}=\boldsymbol{u}\times\boldsymbol{H}$
is defined by the part of the force which does depend on $\boldsymbol{u}$
$-$ we may call it the\emph{ dynamic} (magnetic-like) part of the
force.

To prove the statement above, let us assume that $\boldsymbol{F}$,
the force acting on the body in the frame $R$, does not depend on
$\boldsymbol{u}$ (see footnote \footnote{For speed-dependent force fields, in general there is no inertial frame where the force becomes independent of the body velocity. Nonetheless,
this fact does not invalidate the use of (\ref{F}) and, hence, the
results which follow will still hold. The words \emph{static} and
\emph{dynamic}, however, becomes inappropriate in this case, since
now both $\boldsymbol{G}$ as $\boldsymbol{H}$ may depend on the
body velocity (perhaps the terms \emph{electric-like} and \emph{magnetic-like}
are more suitable here).}). Hence, in the frame $R$ the force is already written in Lorentz
form with $\boldsymbol{G}=\boldsymbol{F}$ and $\boldsymbol{H}=0$.
Now we have to prove that the same is true in an arbitrary inertial
frame $R'$. To find the force in the frame $R'$, we can use (\ref{F}).
Notice however that force $\boldsymbol{F}'$ depends on velocity $\boldsymbol{u}$
that the particle has in the frame $R$. However, the observers in
$R'$ do not measure $\boldsymbol{u}$, instead, it is $\boldsymbol{u}'$
that is actually measured. This fact suggests us to eliminate $\boldsymbol{u}$
through the velocity transformation formulæ \cite{Resnick1968,French1968},
\begin{equation}
u_{x}=\frac{u_{x}'+v}{1+\frac{u_{x}'v}{c^{2}}},\quad u_{y}=\frac{u_{y}'/\gamma}{1+\frac{u_{x}'v}{c^{2}}},\quad u_{z}=\frac{u_{z}'/\gamma}{1+\frac{u_{x}'v}{c^{2}}},\label{u}
\end{equation}
in order to rewrite (\ref{F}) in terms of $\boldsymbol{u}'$. Inserting
(\ref{u}) into (\ref{F}) and simplifying, we get the formulæ \begin{subequations}\label{F'}
\begin{align}
F_{x}' & =F_{x}-\gamma v/c^{2}\left(u_{y}'F_{y}+u_{z}'F_{z}\right),\\
F_{y}' & =\gamma F_{y}\left(1+u_{x}'v/c^{2}\right),\\
F_{z}' & =\gamma F_{z}\left(1+u_{x}'v/c^{2}\right).
\end{align}
\end{subequations} Thus, from the part of $\boldsymbol{F}'$ which
does not depend on $\boldsymbol{u}'$ we get the static part of the
force, $\boldsymbol{G}'$, namely,
\begin{equation}
G_{x}'=F_{x},\quad\ G_{y}'=\gamma F_{y},\quad G_{z}'=\gamma F_{z},\label{G}
\end{equation}
and, from that part of $\boldsymbol{F}'$ which does depend on $\boldsymbol{u}'$,
we get the dynamic part of the force, $\boldsymbol{M}'$,
\begin{subequations}\label{M}
\begin{align}
M_{x}' & = -\gamma v/c^{2}  \left( u_{y}'F_{y} + u_{z}'F_{z} \right), \\
M_{y}' & = \gamma  v /c^{2} \left( u_{x}' F_{y} \right), \\
M_{z}' & = \gamma  v /c^{2} \left( u_{x}' F_{z} \right).
\end{align}
\end{subequations}
Now it is just a matter of fact that (\ref{M}) can be written as
the vector product $\boldsymbol{M}'=\boldsymbol{u}'\times\boldsymbol{H}'$,
with
\begin{equation}
\boldsymbol{H}'=-\boldsymbol{v}/c^{2}\times\boldsymbol{G}'.\label{H}
\end{equation}
Thus, the force $\boldsymbol{F}'$ can be written in the Lorentz form
(\ref{Lorentz}) with respect to the frame $R'$ as well. Moreover, since the
frame $R'$ is quite arbitrary, we had proved that in any inertial
frame every physically acceptable force field can be written in the
Lorentz form\footnote{In a tensor notation, this means that the four-force can always be written as $f^{\alpha}=w_{\alpha}F^{\alpha \beta}$,
where $w_{\alpha}$ denotes the (covariant) components of the four-velocity and $F^{{\alpha \beta}}$ is a suitable anti-symmetric tensor. The correspondent decomposition  $f^{\alpha}=g^{\alpha}+m^{\alpha}$ of the total four-force into the static and dynamic ones is provided by $g^{\alpha}=w_0F^{\alpha 0}$ and $m^{\alpha}=w_{\beta} F^{\alpha \beta} - w_0F^{\alpha 0}$. This means that the spatial part of the four-force can be written as $\bm{f}=\bm{g}+\bm{w}\times\bm{h}$, with $\bm{g}=w_0\left(F^{10}\hat{\boldsymbol{x}} + F^{20}\hat{\boldsymbol{y}} + F^{30}\hat{\boldsymbol{z}}\right)$ and $\bm{h}= F^{23}\hat{\boldsymbol{x}} + F^{31}\hat{\boldsymbol{y}} + F^{12}\hat{\boldsymbol{z}}$.}. This can be also proved considering another inertial frame $R''$ moving with respect to $R'$ with a velocity $\boldsymbol{w}=w\hat{\boldsymbol{x}}$.
Supposing that in $R'$ the force acting on the body is $\boldsymbol{F}'=\boldsymbol{G}'+\boldsymbol{u}'\times\boldsymbol{H}'$,
then, repeating the above procedure, we can show that in $R''$ the
force is still given by $\boldsymbol{F}''=\boldsymbol{G}''+\boldsymbol{u}''\times\boldsymbol{H}''$,
with the static and dynamic forces in each frame related with themselves
by the formulæ \begin{subequations}
\begin{flalign}
G_{x}''=G_{x}', &  & H_{x}''=H_{x}',\\
G_{y}''=\gamma_{w}\left(G_{y}'-wH_{z}'\right), &  & H_{y}''=\gamma_{w}\left(H_{y}'+\tfrac{w}{c^{2}}G_{z}'\right),\\
G_{z}''=\gamma_{w}\left(G_{z}'+wH_{y}'\right), &  & H_{z}''=\gamma_{w}\left(H_{z}'-\tfrac{w}{c^{2}}G_{y}'\right),
\end{flalign}
\end{subequations} where $\gamma_{w}=1/\sqrt{1-w^{2}/c^{2}}$. The
most known example of such a force is the electromagnetic one. In
this case we have $\boldsymbol{G}=q\boldsymbol{E}$ and $\boldsymbol{M}=q\boldsymbol{u}\times\boldsymbol{B}$,
with $q$ denoting the electric charge, $\boldsymbol{E}$ the electric
field and $\boldsymbol{B}$ the magnetic field, respectively. Another
example is the gravitational force in the approximation where Newton's
law is valid (\emph{i.e.,} in a flat spacetime background). In this
case, we have $\boldsymbol{G}=m\boldsymbol{g}$ and $\boldsymbol{M}=m\boldsymbol{u}\times\boldsymbol{h}$,
where $\boldsymbol{g}$ is the static gravitational field and $\boldsymbol{h}$
the dynamic gravitational field $-$ the gravitational analogue of
the magnetic field. The consequences of this covariant theory of gravitation were recently discussed in \cite{Vieira2016}.

\section{Solution of the submarine paradox in a flat spacetime\label{Sec: Flat}}

Let us then analyze the submarine paradox but considering, by now,
only the special theory of relativity. This means that in this section
the gravitational interaction will be regarded as an ordinary force
field in a flat spacetime (in the same way as, for instance, the electromagnetic
interactions are usually treated in special relativity). We shall
assume therefore that the gravitational force between the Earth and
a given particle is determined by Newton's law,
\begin{equation}
\boldsymbol{F}=-\left(GMm/r^{2}\right)\hat{\boldsymbol{r}},\label{Newton}
\end{equation}
when the Earth is at rest, no matter what is the motion of the particle.
In (\ref{Newton}), $M$ is the mass of Earth, $m$ the mass of the
particle, $\boldsymbol{r}$ is the distance vector from the Earth
to the particle position, $G$ the Newton constant. In the present
case, we shall consider actually only the constant gravitational force
$\boldsymbol{F}=-mg\hat{\boldsymbol{z}}$ on the surface of Earth,
where $g$ is the acceleration of gravity. Finally, we shall also
consider that the (inertial and gravitational) mass is an invariant
quantity $-$ which is, of course, the most logical way to proceed
in order to avoid misunderstandings\footnote{See \cite{Adler1987, Okun1989, Okun1989B, Okun2009, Jammer2000}
for discussions about the concept of mass in relativity.}.

Before analyze the original formulation of Supplee's paradox, let
us consider first a slightly modified version in which no acceleration
is involved. This is obtained supposing a submarine moving with velocity
$\boldsymbol{v}=v\hat{\boldsymbol{x}}$ in the standing water of the
ocean and letting its density be adjusted by the observers at rest
within the ocean (frame $R$) in such a way that the submarine remains
in equilibrium in this frame. From the Archimedes principle this means
that the submarine density must be adjusted to be the same as the
water density when both are measured by the frame $R$. The paradox
situation arises because it seems, at first sight, that the submariners
(frame $R'$) would conclude that the submarine should float, since in
this frame the submarine density happens to be lesser than that of
the moving water, thanks to the Lorentz length effects. We shall see,
however, that this apparent paradox is due to an incorrect use of
the ordinary Archimedes principle: if special relativity is correctly
employed, we shall obtain that the submarine neither sinks nor floats in both frames (see figs. 1 and 2).

A straightforward confirmation of this result could be given directly
from the transformation formulæ (\ref{F}) or (\ref{F'}). In fact,
since in the frame $R$ the total force acting on the submarine is
null, the same will be true in the frame $R'$, so that the submarine
cannot accelerate in neither frame. However, in order to provide a
physical explanation of the problem, a more detailed exposition is
necessary.

Let us begin our analysis in the frame $R$. Here, the water of the
ocean is at rest, while the submarine has velocity $\boldsymbol{v}=v\hat{\boldsymbol{x}}$.
The submarine is subject to two forces: a small constant gravitational
(weight) force, $\boldsymbol{W}=-mg\hat{\boldsymbol{z}}=-\rho_{\mathrm{s}}V_{\mathrm{s}}g\hat{\boldsymbol{z}}$,
and the Archimedes (buoyancy) force, $\boldsymbol{A}$. Archimedes
force is a response from the water to the action of gravity: a gradient
of pressure arises in order to keep its static equilibrium. The gradient
of pressure present in a fluid suited in a gravitational field equals
the density of that gravitational force, $\boldsymbol{\nabla}p=\boldsymbol{f}=-\rho_{\mathrm{w}}g\hat{\boldsymbol{z}}$.
Hence, Archimedes force acting on the submarine can be found integrating
$-\boldsymbol{\nabla}p$ over the submarine volume:
\begin{equation}
\boldsymbol{A}=\int_{V_{\mathrm{s}}}\left(-\boldsymbol{\nabla}p\right)dV=\int_{V_{\mathrm{s}}}\rho_{\mathrm{w}}g\hat{\boldsymbol{z}}dV=\rho_{\mathrm{w}}V_{\mathrm{s}}g\hat{\boldsymbol{z}}.
\end{equation}
Thus we can see that the total force acting on the submarine will
be null if we set $\rho_{\mathrm{s}}=\rho_{\mathrm{w}}=\rho$.

In the proper submarine frame, $R'$, on the other hand, the water
is moving with the velocity $\boldsymbol{u}_{\mathrm{w}}'=-v\hat{\boldsymbol{x}}$.
Lorentz length effects imply that water's density increases, while
the submarine density decreases by the same factor:
\begin{equation}
\rho_{\mathrm{w}}'=\gamma\rho,\quad\rho_{\mathrm{s}}'=\rho/\gamma.\label{rho'}
\end{equation}
In this frame, the submarine is also subject to the gravitational
and Archimedes forces. The gravitational force, however, is not given
by Newton's law anymore, neither can the Archimedes force be deduced
directly from the usual Archimedes principle. The gravitational force
acting on the submarine should be found through (\ref{F'}), remembering
that in $R'$ the submarine velocity $\boldsymbol{u}_{\mathrm{s}}'$
is zero:
\begin{figure}\label{fig1}
\begin{centering}
\includegraphics[width=8cm]{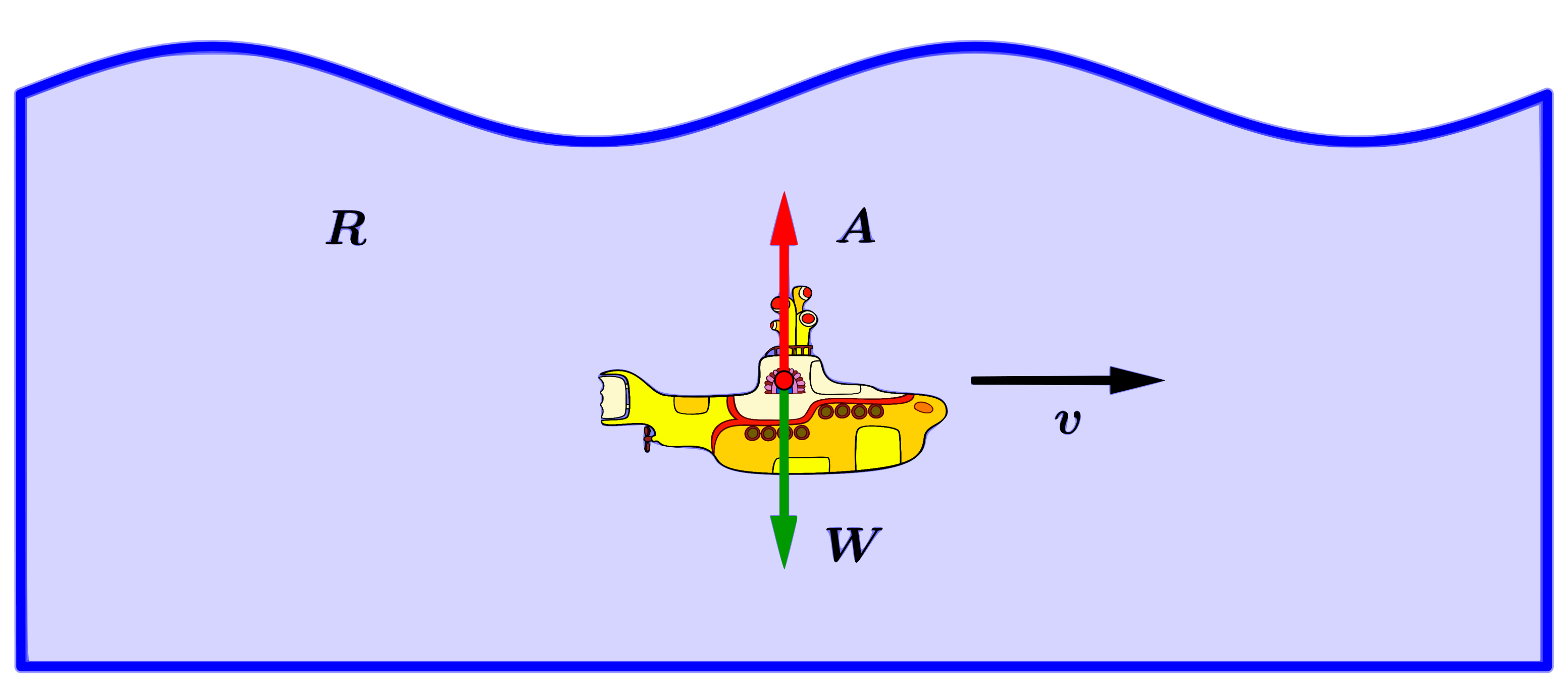}
\par\end{centering}
\label{Fig1}\caption{\emph{Submarine paradox in the frame $R$}. A submarine moves
with velocity $\boldsymbol{u}_{\mathrm{s}}=v\hat{\boldsymbol{x}}$ in a standing
ocean. The density of the moving submarine is adjusted to the water
density, so that the Archimedes force $\boldsymbol{A}$ cancels the gravitational force $\boldsymbol{W}$ and the submarine remains in equilibrium.}
\end{figure}
\begin{equation}
\boldsymbol{W}'=\gamma\boldsymbol{W}=-\gamma mg\hat{\boldsymbol{z}}=-\gamma\rho V_{\mathrm{s}}g\hat{\boldsymbol{z}}.
\end{equation}

On the other hand, to find the Archimedes force in the frame $R'$,
we need to know first what happens with the gradient of pressure on
the water with respect to this frame. Of course, the gradient of pressure can
also be regarded here as a response from the water to the gravitational
force of the Earth, but now we should realize that both the water
as the Earth move with respect to $R'$ with the velocity $\boldsymbol{u}_{\mathrm{wat}}'=-v\hat{\boldsymbol{x}}$.
Hence, the water will be subject to both a static (electric-like)
gravitational force as well as to a dynamic (magnetic-like) one. A unit
volume of water in the frame $R$ is subject to the gravitational
force $\boldsymbol{F}=-\rho g\hat{\boldsymbol{z}}$ and, from (\ref{G})
and (\ref{M}), it follows that the static and dynamic gravitational
forces acting on this element of volume, as measured in the frame
$R'$, will be, respectively,
\begin{equation}
\boldsymbol{G}'=-\gamma\rho g\hat{\boldsymbol{z}},\quad\boldsymbol{M}'=\gamma\left(v^{2}/c^{2}\right)\rho g\hat{\boldsymbol{z}}.\label{G'}
\end{equation}
Notice that, according to (\ref{M}) and (\ref{H}) we can write the
dynamic force as $\boldsymbol{M}'=\boldsymbol{u}_{\mathrm{w}}'\times\boldsymbol{H}'$,
where $\boldsymbol{H}'=-\boldsymbol{v}/c^{2}\times\boldsymbol{G}'$.
From (\ref{Lorentz}) we get, therefore, the total force acting on
that element of volume:
\begin{equation}
\boldsymbol{F}'=-\gamma\rho g\hat{\boldsymbol{z}}+\gamma\left(v^{2}/c^{2}\right)\rho g\hat{\boldsymbol{z}}=-\rho g\hat{\boldsymbol{z}}/\gamma.\label{FW}
\end{equation}
However, this quantity of water no longer occupies a unit volume in
the frame $R'$. In fact, it is contracted by a factor of $1/\gamma$,
so that, in order to get the force per unit volume in the frame $R'$,
we need further to divide (\ref{FW}) by this factor. Whence, we get,
\begin{equation}
\boldsymbol{f}'=\boldsymbol{\nabla}'p'=-\rho g\hat{\boldsymbol{z}}.\label{FP}
\end{equation}
The conclusion is that the gradient of pressure is not proportional
to the higher water's density, as one could naïvely think: rather, it
is an invariant quantity, which could be anticipated already from
the fact that pressure is a scalar and from $\partial_{z}'=\partial_{z}$.
This is why the ordinary Archimedes principle cannot be applied in
$R'$.
\begin{figure}\label{fig2}
\begin{centering}
\includegraphics[width=8cm]{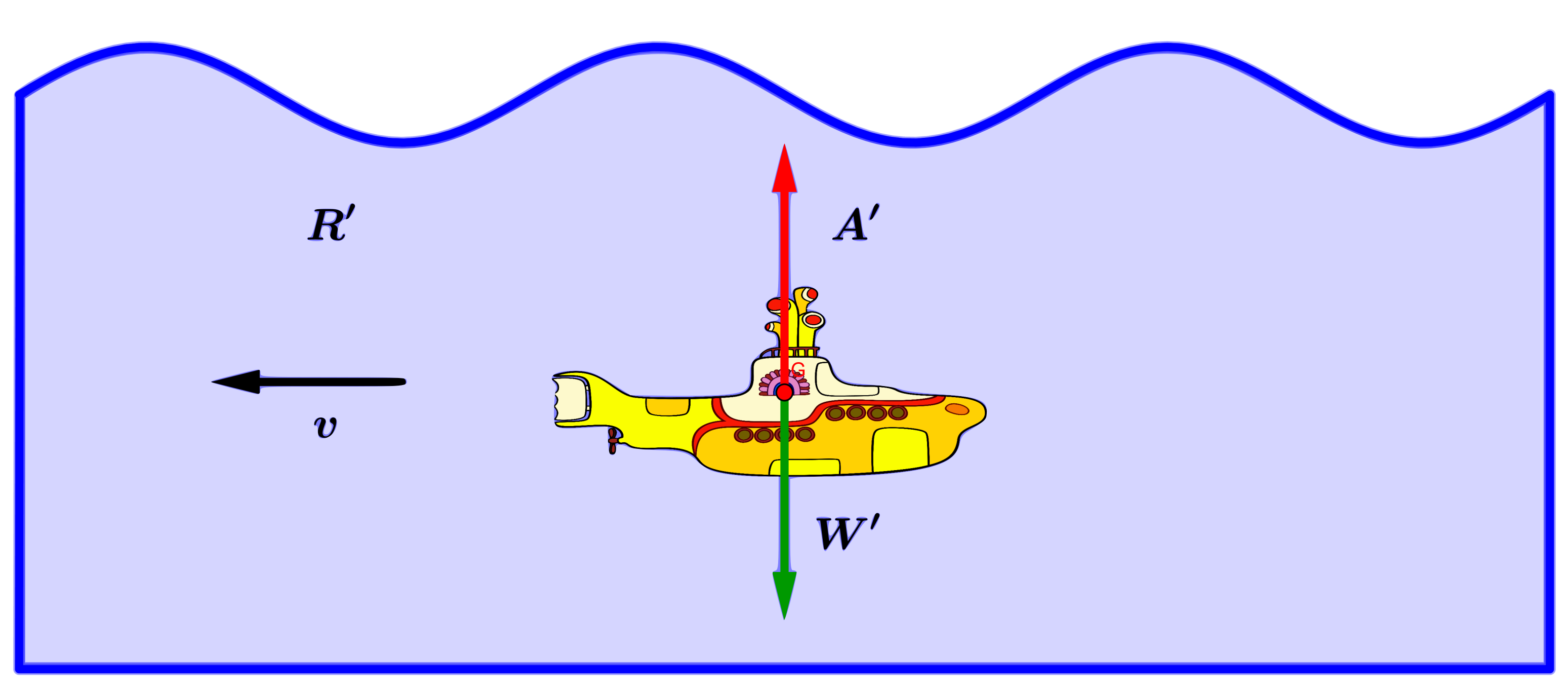}
\par\end{centering}
\label{Fig2}\caption{\emph{Submarine paradox in the frame $R'$}. Here the submarine is at rest while the water moves with the velocity $\boldsymbol{u}'_{\mathrm{w}}=-v\hat{\boldsymbol{x}}$.
Due to Lorentz length effects, the submarine density becomes lower
than water's density. Nevertheless, a relativistic version of the Archimedes principle ensures that the submarine still remains in equilibrium.
The dynamic (magnetic-like) gravitational force that
the moving Earth exerts on the moving water contributes significantly
to this result.}
\end{figure}

Integrating $-\boldsymbol{\nabla}'p'$ over the submarine
volume, we get the Archimedes force acting on it:
\begin{equation}
\boldsymbol{A}'=\int_{V_{\mathrm{s}}'}\left(-\boldsymbol{\nabla}'p'\right)dV'=\int_{V_{\mathrm{s}}'}\rho g\hat{\boldsymbol{z}}dV'=\rho V_{\mathrm{s}}'g\hat{\boldsymbol{z}}.
\end{equation}
Finally, since $V_{\mathrm{s}}'=\gamma V_{\mathrm{s}}$, we get,
\begin{equation}
\boldsymbol{A}'=\gamma\rho V_{\mathrm{s}}g\hat{\boldsymbol{z}}=-\boldsymbol{W}'.\label{A}
\end{equation}
Therefore, the Archimedes force intensity equals the weight of the
submarine and, thus, it neither floats nor sinks in the frame $R'$
as well $-$ the submarine remains in equilibrium in both frames.

We would like to highlight that the Archimedes force can also be obtained
from a \emph{relativistic Archimedes principle}\footnote{We also point out that such a relativistic Archimedes principle can be important for the analysis of the equilibrium of (very fast) rotating stars, since the relative motion between their layers would prevent us from using the usual Archimedes principle.}. Remember that the
original formulation of the Archimedes principle states that the intensity
of the Archimedes force equals the weight of water displaced by the
submersed body. This principle is still valid in the frame $R'$,
but here we must distinguish between the \emph{static weight} of the
mass displaced by the body, defined as the displaced mass times the
static (electric-like) gravitational field $\boldsymbol{W}'=\rho_{\mathrm{w}}'V_{\mathrm{b}}'\boldsymbol{g}'$,
and its \emph{dynamic weight}, which is given by the current of mass
times the dynamic (magnetic-like) gravitational field, $\boldsymbol{M}'=V_{\mathrm{b}}'\boldsymbol{j}_{\mathrm{w}}'\times\boldsymbol{h}'$,
where $\boldsymbol{j}_{\mathrm{w}}'=\rho_{\mathrm{w}}'\boldsymbol{u}_{\mathrm{w}}'$
and $\boldsymbol{h}'=-\boldsymbol{v}/c^{2}\times\boldsymbol{g}'$.
Therefore, we get for the static and dynamic weight of the submarine,
respectively,
\begin{equation}
\boldsymbol{W}'=-\gamma\rho_{\mathrm{w}}'V_{\mathrm{s}}'g\hat{\boldsymbol{z}},\quad\boldsymbol{M}'=\gamma\left(v^{2}/c^{2}\right)\rho_{\mathrm{w}}'V_{\mathrm{s}}'g\hat{\boldsymbol{z}}.
\end{equation}
Notice that the dynamic weight is contrary to the static weight
and hence it can be thought as a ``negative weight'' due to the
repulsive dynamic gravitational force between the Earth and the ocean
water. The sum of these two terms (with the opposite signs) provides,
of course, the Archimedes force (\ref{A}).

Finally, let us consider the original formulation of Supplee's paradox.
In this case the density of the submarine is adjusted to the water
density when both of them are at rest (let $m$ be the submarine mass
and $V_{0}$ its proper volume, so that its proper density is $\rho_{0}=m/V_{0}$).
If the submarine is put to move with a velocity $\boldsymbol{v}=v\hat{\boldsymbol{x}}$,
the gravitational force acting on it will still be $\boldsymbol{W}=-mg\hat{\boldsymbol{z}}=-\rho_{0}V_{0}g\hat{\boldsymbol{z}}$,
since the gravitational field is just static in the frame $R$. To
evaluate the Archimedes force we should realize that now the submarine
volume is contracted to $V_{\mathrm{s}}=V_{0}/\gamma$, and then,
from the Archimedes principle, we get that $\boldsymbol{A}=\rho_{0}V_{0}g\hat{\boldsymbol{z}}/\gamma$.
Thus, the total force acting on the submarine is
\begin{equation}
\boldsymbol{F}=-\rho_{0}V_{0}g\left(1-1/\gamma\right)\hat{\boldsymbol{z}}.\label{FR}
\end{equation}

On the other hand, in the frame $R'$ (the inertial frame that is
instantaneously at rest with respect to the submarine at at $t=t'=0$), the
weight force acting on the submarine will be, according to (\ref{F'}),
$\boldsymbol{W}'=-\gamma\rho_{0}V_{0}g\hat{\boldsymbol{z}}$. Notice
that there is no dynamic (magnetic-like) force here again, since the
submarine is at rest on $R'$ in this instant of time. There is, however, a dynamic force between the moving Earth and the moving ocean.
As we have seen, these gravitomagnetic forces combined imply that the
gradient of water's pressure observed in the frame $R'$ is the same
as that measured in $R$.
Thus, Archimedes force will be given just by $\boldsymbol{A}'=\rho_{0}V_{0}g\hat{\boldsymbol{z}}$, from which we get the total force acting on the submarine in the frame $R'$:
\begin{equation}
\boldsymbol{F}'=-\rho_{0}V_{0}g\left(\gamma-1\right)\hat{\boldsymbol{z}}=\gamma\boldsymbol{F}.\label{FR'}
\end{equation}

The conclusion is that in both frames the submarine will sink. Notice
further that is not necessary to employ accelerated frames neither
general relativity in order to study the submarine behavior. We can
do that, of course, but then we should take care of the geometric
effects that arise in non-inertial frames, as already discussed by
Supple, Matsas and Jonsson \cite{Supplee1989,Matsas2003,Jonsson2006}.

\section{Spacetime curvature in the vicinity of Earth and the implicated speed-dependent
gravitational force\label{Sec: Force}}

According to Einstein's theory of gravitation, gravity is not a force
but just an effect of the spacetime curvature \cite{Taylor2000,Misner1973,Landau1980}.
In other words, Einstein's theory implies that there is no gravitational
field in an exactly flat spacetime whatsoever. Of course, the Newtonian
description of gravity as a force field in a flat spacetime can be
justified as being a very good approximation in the vicinity of the Earth,
which is due to the very small curvature of spacetime there $-$ so
small that we can only measure its effects with the most precise instruments
available. The former approach presented in the last section assumes
that this is indeed the case, hence, it should be treated as a first
approximation for the problem.

Nevertheless, in order to get a better approximation for the gravitational
phenomenon on the surface of Earth, the curvature of spacetime need
to be taken into account. Since any physical measurement of distances
performed on the surface of the Earth does not reveal any discordance
with the Euclidean geometry, it is enough to consider here a \emph{flat
space plus a curved time}. This means that, {in this approximation, the metric on the proximity of the Earth can be written in the form,
\begin{equation}
\upd s^{2}=-f\left(z\right)c^{2}\upd t^{2}+\upd x^{2}+\upd y^{2}+ \upd z^{2},\label{Metric}
\end{equation}
(we assume that the Earth is large enough so that its surface can be
approximated by the horizontal plane), where the function $f\left(z\right)$
is to be determined. In order to do so, we shall proceed as follows:
first remember that in Einstein's theory, the world-line of a particle
freely falling in a gravitational field is a geodesic, which is determined
by the equations \cite{Misner1973,Landau1980}
\begin{equation}
\upd u^{\alpha}/\upd \tau+\varGamma_{\beta\gamma}^{\alpha}u^{\beta}u^{\gamma}=0,\label{Geodesic}
\end{equation}
where $u^{\alpha}$ are the components of the particle four-velocity,
$\tau$ is its proper time and $\varGamma_{\beta\gamma}^{\alpha}$
are the Christoffel symbols, which are obtained from the metric through
the formula \cite{Misner1973,Landau1980}
\begin{equation}
\varGamma_{\beta\gamma}^{\alpha}=\tfrac{1}{2}g^{\alpha\delta}\left(\partial_{\beta}g_{\gamma\delta}+\partial_{\gamma}g_{\beta\delta}-\partial_{\delta}g_{\beta\gamma}\right).\label{Connection}
\end{equation}

The geodesic equation agrees with the fact that the particle four-acceleration
$A^{\alpha}=\upd u^{\alpha}/\upd\tau+\varGamma_{\beta\gamma}^{\alpha}u^{\beta}u^{\gamma}$
must be zero in the proper co-moving frame of the particle, since
Einstein's equivalence principle states that the particle does not
feel any effect of gravity as it freely falls in the gravitational
field. We may call the fist term $a^{\alpha}=\upd u^{\alpha}/\upd\tau$ in the formula above as the
\emph{kinematic four-acceleration} of the particle,
while the second term $\varGamma^{\alpha}=\varGamma_{\beta\gamma}^{\alpha}u^{\beta}u^{\gamma}$
can be called its \emph{geometric four-acceleration} (since it is
present only when curved coordinates are employed to describe the
motion of the particle). Now, consider a fixed observer very close to the particle position
(\emph{e.g.,} both of them in the proximity of the Earth). The difference
between the observer and the particle proper times will be only
due to the relative motion between them. In fact, any effect arising from the metric will cancel, since the metric
will be the same to both the particle and this observer. Therefore, we can write for the particle four-acceleration,
as measured by this static observer,
\begin{equation}
a^{\alpha}=-\varGamma_{\beta\gamma}^{\alpha}u^{\beta}u^{\gamma}.\label{a}
\end{equation}
Hence, if a particle is released from rest near the Earth surface,
in this very instant the components of its (relative)  four-velocity will be
\begin{equation}
u^{0}=c,\quad u^{1}=0,\quad u^{2}=0,\quad u^{3}=0.\label{u4}
\end{equation}
Besides, we know that in this case the acceleration of the particle
is just $\boldsymbol{a}=-g\hat{\boldsymbol{z}}$ and, thus, the components
of its four-acceleration will be as well,
\begin{equation}
a^{0}=0,\quad a^{1}=0,\quad a^{2}=0,\quad a^{3}=-g.\label{g}
\end{equation}

On the other hand, it follow from (\ref{Metric}) and (\ref{Connection})
that the only non-null Christoffel symbols are
\begin{equation}
\varGamma_{03}^{0}=\varGamma_{30}^{0}=\frac{1}{2}\frac{\upd}{\upd z}\log f\left(z\right),\quad\mathrm{and}\quad\varGamma_{00}^{3}=\frac{1}{2}\frac{\upd  f\left(z\right)}{ \upd z}.\label{Conection-F}
\end{equation}
Then, using (\ref{a}), (\ref{u4}), (\ref{g}) and (\ref{Conection-F})
at once, we get the relation
\begin{equation}
a^{3}=-\varGamma_{00}^{3}u^{0}u^{0}=-c^{2}/2\left[\upd f\left(z\right)/ \upd z\right]=-g,
\end{equation}
and, solving this equation, we find that $f(z)=2gz/c^{2}+C$,
where $C$ is the constant of integration. In order to fix $C$, we
may realize that in the absence of the gravitational field (\emph{i.e.},
for $g=0$), the metric should reduce to that of Minkowski,  $\upd s^{2}=-c^{2}\upd t+\upd x^{2}+\upd y^{2}+\upd z^{2}$, which lead us to the value $C=1$. Hence, the spacetime metric in the vicinity of the Earth's surface becomes,
\begin{equation}
\upd s^{2}=-\left(1+2gz/c^{2}\right)c^{2}\upd t^{2}+\upd x^{2}+\upd y^{2}+\upd z^{2}.\label{Metric-Final}
\end{equation}
The corresponding non-null Christoffel symbols reduce to
\begin{equation}
\varGamma_{03}^{0}=\varGamma_{30}^{0}=g/c^{2} \left(1+2gz/c^{2}\right)^{-1},\quad\varGamma_{00}^{3}=g/c^{2}.\label{Connection-Final}
\end{equation}
It can be verified that the metric (\ref{Metric-Final}) satisfies
the Einstein field equations $G_{\alpha\beta}=\kappa T_{\alpha\beta}$, where the only non-null elements of the energy-momentum tensor $T_{\alpha\beta}$ are
\begin{equation}
 T_{11}=T_{22}=-g^{2}/\kappa c^{4}\left(1+2gz/c^{2}\right)^{2}.
\end{equation}
As we can see, the values of $T_{11}$ and $T_{22}$ are very small
in the proximity of Earth, so that we get an almost vacuum solution there which is quite reasonable in our approximation. Nevertheless,
their exact values provide an exact solution of Einstein's field equations
for a universe described by the metric (\ref{Metric-Final}).
The negative values of $T_{11}$ and $T_{22}$ can be interpreted as negative vacuum pressures that ensure the static behavior of such a universe.

Now, let us see what should be the acceleration of a particle which moves with a given (instantaneous) velocity $\boldsymbol{u}$ in the gravitational field of the Earth. In this case, its four-velocity becomes,
\begin{equation}
u^{0}=\gamma_{u}c,\quad u^{1}=\gamma_{u}u_{x},\quad u^{2}=\gamma_{u}u_{y},\quad u^{3}=\gamma_{u}u_{z},\label{u4P}
\end{equation}
 However, we cannot assume that the acceleration of the particle is
directed along the $z$ direction anymore, since
the theory of relativity shows us that acceleration and force are
not parallel one to the other, except when the velocity is parallel
or orthogonal to the force. In fact, the relationship between force
and acceleration is \cite{Resnick1968,French1968}
\begin{equation}
\boldsymbol{F}=m\left[\gamma_{u}\boldsymbol{a}+\gamma_{u}^{3}\left(\boldsymbol{a}\cdot\boldsymbol{u}\right)\boldsymbol{u}/c^{2}\right].\label{FA}
\end{equation}
Hence, the particle four-acceleration must be written as
\begin{equation}
a^{0}=\gamma_{u}^{4}\left(\frac{\boldsymbol{a}\cdot\boldsymbol{u}}{c}\right),\quad a^{1,2,3}=\gamma_{u}^{2}a_{x,y,z}+\gamma_{u}^{4}\left(\frac{\boldsymbol{a}\cdot\boldsymbol{u}}{c^{2}}\right)u_{x,y,z}.\label{a4m}
\end{equation}
On the other hand, (\ref{a}), (\ref{Connection-Final}) and (\ref{u4P})
give,
\begin{equation}
a^{0}=-\frac{2g\gamma_{u}^{2}u_{z}/c}{\left(1+2gz/c^{2}\right)},\quad a^{1}=0,\quad a^{2}=0,\quad a^{3}=-g\gamma_{u}^{2}.\label{ag}
\end{equation}
Comparing (\ref{a4m}) with (\ref{ag}) we obtain the acceleration
of the particle in the weak gravitational field of Earth:
\begin{equation}
a_{x}=\frac{u_{x}u_{z}}{c^{2}}g,\quad
a_{y}=\frac{u_{y}u_{z}}{c^{2}}g,\quad a_{z}=-g\left(1-\tfrac{u_{z}^{2}}{c^{2}}\right).\label{Au}
\end{equation}
Notice that the particle acceleration will be directed along the $z$
axis only if $u_{z}=0$ or if $u_{x}=u_{y}=0$. In the fist case,
the acceleration is $\boldsymbol{a}=-g\hat{\boldsymbol{z}}$, while
in the second case we have $\boldsymbol{a}=-g/\gamma_{u}^{2}\hat{\boldsymbol{z}}$.

Finally, inserting (\ref{Au}) into (\ref{FA}) and simplifying, we
get the gravitational force acting on the moving particle:
\begin{equation}
\boldsymbol{F}(u)=-\gamma_{u}mg\hat{\boldsymbol{z}}.\label{NewtonV}
\end{equation}

We conclude therefore that the spacetime curvature in the proximity
of Earth implies a \emph{speed-dependent gravitational force}. The
gravitational force increases with the particle speed.

Notwithstanding the curved spacetime we have here, we may realize
that the Lorentz transformations can still be employed, as long as
the frame $R'$ moves with respect to the frame $R$ in a direction parallel to Earth's surface.
The only effects differing from those obtained in an exactly flat spacetime are those on which events are to be compared in different heights, since, according to the metric (\ref{Metric-Final}), the higher the position is, the faster the time will pass. In fact, if $\upd t$ is a given interval of time measured by a clock at the surface of the Earth, then an identical clock situated at the height $z$ will mark the time
$\upd t_{z}=\sqrt{1+2gz/c^{2}}\upd t$, which is greater than $\upd t$. Correspondingly, if a light ray is emitted from the height $z$ towards the surface of Earth, a blue-shift effect will take place.

\section{Solution of the submarine paradox in the curved spacetime of Earth\label{Sec: Curved}}

Finally, let us analyze what change in the description of the submarine
paradox, when the effects of the tiny spacetime curvature on the Earth
surface are taken into account. It is sufficient in this case to consider
a gravitational force law given by (\ref{NewtonV}), instead of
Newton's law (\ref{Newton}).

Let us first consider that version of the paradox on which the submarine
does not accelerate. In this case, where the spacetime is weakly curved,
the gravitational force acting on the submarine with respect to the frame $R$, will be
$\boldsymbol{W}=-\gamma mg\hat{\boldsymbol{z}}=-\gamma\rho_{\mathrm{s}}V_{\mathrm{s}}g\hat{\boldsymbol{z}}$,
since the submarine moves with the velocity $\boldsymbol{v}=v\hat{\boldsymbol{x}}$
and we are using the force law (\ref{NewtonV}). In order for the submarine
to stay in equilibrium underwater, its density should now be adjusted (by the observers at rest within the water) to
$\rho_{\mathrm{s}}=\rho_{\mathrm{w}}/\gamma$, so that the intensity of Archimedes force $\boldsymbol{A}$ equals
the intensity of the weight force $\boldsymbol{W}$. In the frame
$R'$, of course, the same will be true. In fact, the gravitational
force acting on the submarine can be found through (\ref{F'}), and
it is given by $\boldsymbol{W}'=-\gamma^{2}mg\hat{\boldsymbol{z}}=-\gamma^{2}\rho_{\mathrm{s}}V_{\mathrm{s}}g\hat{\boldsymbol{z}}$.
The density of the water in the frame $R'$ is still given by (\ref{rho'})
and, hence, the gradient of pressure in the frame $R'$ remains the
same as that measured in the frame $R$. Thus, Archimedes force becomes,
$\boldsymbol{A}'=\gamma\rho_{\mathrm{w}}V_{\mathrm{s}}g\hat{\boldsymbol{z}}=\gamma^{2}\rho_{\mathrm{s}}V_{\mathrm{s}}g\hat{\boldsymbol{z}}$,
from which we can see that in frame $R'$ the submarine will remain
in equilibrium as well.

For the original formulation of Supplee's paradox, we get a similar
explanation. Here the submarine density $\rho_{0}$ is matched with
the water density when both of them are at rest. In the frame $R$, the submarine
moves with the velocity $\boldsymbol{v}=v\hat{\boldsymbol{x}}$ and
gravitational force acting on it is $\boldsymbol{W}=-\gamma mg\hat{\boldsymbol{z}}=-\gamma\rho_{0}V_{\mathrm{0}}g\hat{\boldsymbol{z}}$.
The Archimedes force is $\boldsymbol{A}=\rho_{0}V_{\mathrm{s}}g\hat{\boldsymbol{z}}=\rho_{0}V_{0}g\hat{\boldsymbol{z}}/\gamma$
and, thus, the total force acting on the submarine is $\boldsymbol{F}=-\rho_{0}V_{\mathrm{0}}g\hat{\boldsymbol{z}}\left(\gamma-1/\gamma\right)$.
In the frame $R'$ the total force acting on the submarine can be found through (\ref{F'}), and then we get
$\boldsymbol{F}'=-\gamma\rho_{0}V_{\mathrm{0}}g\hat{\boldsymbol{z}}\left(\gamma-1/\gamma\right)$.

These expressions for the total force acting on the submarine agree with those obtained
by Supplee and Matsas \cite{Supplee1989,Matsas2003}. It should be
mentioned, however, that the argument presented by Supplee in his
first explanation of the paradox, concerning the gravitational force
between moving bodies, cannot be justified. In fact, Supplee assumed
in \cite{Supplee1989} that Newton's law (\ref{Newton}) is still
valid, even when the interacting bodies are in motion. However he
also assumed (as many others do, see \cite{Adler1987,Okun1989,Okun1989B,Jammer2000,Okun2009})
that the gravitational mass should be determined by Einstein's formula
$E=mc^{2}$, which leads to a speed-dependent mass: $m=\gamma_{u}m_{0}$,
where $m_{0}$ is the so called ``rest mass'' and $\bm{u}$ is the
velocity of the body. This, by its turn, leads to the same speed-dependent
force (\ref{NewtonV}), as deduced by us in the previous section. Notice, however, that the relation between mass and energy expressed by Einstein's
formula above only holds when the momentum of the body is null $-$
in fact the correct expression is $E=\sqrt{m^{2}c^{4}+p^{2}c^{2}}$.
Besides, we must remember that in Einstein's theory of gravitation
the source of gravitational interaction is energy-momentum tensor,
not the energy alone. Therefore, it seems only a coincidental fact
that the speed-dependent mass considered by Supplee and others gives
the same speed-dependent force (\ref{NewtonV}), as deduced from general relativity in the approximation considered here.

\begin{acknowledgments}
I kindly thank Professor G. E. A. Matsas for the discussions we had about this approach to the submarine paradox.
\end{acknowledgments}
\bibliography{references}

\begin{thebibliography}{15}%
\makeatletter
\providecommand \@ifxundefined [1]{%
 \@ifx{#1\undefined}
}%
\providecommand \@ifnum [1]{%
 \ifnum #1\expandafter \@firstoftwo
 \else \expandafter \@secondoftwo
 \fi
}%
\providecommand \@ifx [1]{%
 \ifx #1\expandafter \@firstoftwo
 \else \expandafter \@secondoftwo
 \fi
}%
\providecommand \natexlab [1]{#1}%
\providecommand \enquote  [1]{``#1''}%
\providecommand \bibnamefont  [1]{#1}%
\providecommand \bibfnamefont [1]{#1}%
\providecommand \citenamefont [1]{#1}%
\providecommand \href@noop [0]{\@secondoftwo}%
\providecommand \href [0]{\begingroup \@sanitize@url \@href}%
\providecommand \@href[1]{\@@startlink{#1}\@@href}%
\providecommand \@@href[1]{\endgroup#1\@@endlink}%
\providecommand \@sanitize@url [0]{\catcode `\\12\catcode `\$12\catcode
  `\&12\catcode `\#12\catcode `\^12\catcode `\_12\catcode `\%12\relax}%
\providecommand \@@startlink[1]{}%
\providecommand \@@endlink[0]{}%
\providecommand \url  [0]{\begingroup\@sanitize@url \@url }%
\providecommand \@url [1]{\endgroup\@href {#1}{\urlprefix }}%
\providecommand \urlprefix  [0]{URL }%
\providecommand \Eprint [0]{\href }%
\providecommand \doibase [0]{http://dx.doi.org/}%
\providecommand \selectlanguage [0]{\@gobble}%
\providecommand \bibinfo  [0]{\@secondoftwo}%
\providecommand \bibfield  [0]{\@secondoftwo}%
\providecommand \translation [1]{[#1]}%
\providecommand \BibitemOpen [0]{}%
\providecommand \bibitemStop [0]{}%
\providecommand \bibitemNoStop [0]{.\EOS\space}%
\providecommand \EOS [0]{\spacefactor3000\relax}%
\providecommand \BibitemShut  [1]{\csname bibitem#1\endcsname}%
\let\auto@bib@innerbib\@empty
\bibitem [{\citenamefont {Supplee}(1989)}]{Supplee1989}%
  \BibitemOpen
  \bibfield  {author} {\bibinfo {author} {\bibfnamefont {J~M}\ \bibnamefont
  {Supplee}},\ }\bibfield  {title} {\enquote {\bibinfo {title} {Relativistic
  buoyancy},}\ }\href@noop {} {\bibfield  {journal} {\bibinfo  {journal} {Am.
  J. Phys.}\ }\textbf {\bibinfo {volume} {57}},\ \bibinfo {pages} {75}
  (\bibinfo {year} {1989})}\BibitemShut {NoStop}%
\bibitem [{\citenamefont {Matsas}(2003)}]{Matsas2003}%
  \BibitemOpen
  \bibfield  {author} {\bibinfo {author} {\bibfnamefont {G~E~A}\ \bibnamefont
  {Matsas}},\ }\bibfield  {title} {\enquote {\bibinfo {title} {Relativistic
  archimedes law for fast moving bodies and the general-relativistic resolution
  of the `submarine paradox'},}\ }\href@noop {} {\bibfield  {journal} {\bibinfo
   {journal} {Phys. Rev. D}\ }\textbf {\bibinfo {volume} {68}},\ \bibinfo
  {pages} {027701} (\bibinfo {year} {2003})},\ \Eprint
  {http://arxiv.org/abs/arXiv:gr-qc/0305106} {arXiv:gr-qc/0305106} \BibitemShut
  {NoStop}%
\bibitem [{\citenamefont {Jonsson}(2006)}]{Jonsson2006}%
  \BibitemOpen
  \bibfield  {author} {\bibinfo {author} {\bibfnamefont {R~M}\ \bibnamefont
  {Jonsson}},\ }\bibfield  {title} {\enquote {\bibinfo {title} {An intuitive
  approach to inertial forces and the centrifugal force paradox in general
  relativity},}\ }\href@noop {} {\bibfield  {journal} {\bibinfo  {journal} {Am.
  J. Phys.}\ }\textbf {\bibinfo {volume} {74}},\ \bibinfo {pages} {905}
  (\bibinfo {year} {2006})},\ \Eprint {http://arxiv.org/abs/arXiv:0708.2488}
  {arXiv:0708.2488} \BibitemShut {NoStop}%
\bibitem [{\citenamefont {Taylor}\ and\ \citenamefont
  {Wheeler}(2000)}]{Taylor2000}%
  \BibitemOpen
  \bibfield  {author} {\bibinfo {author} {\bibfnamefont {E~F}\ \bibnamefont
  {Taylor}}\ and\ \bibinfo {author} {\bibfnamefont {J~A}\ \bibnamefont
  {Wheeler}},\ }\href@noop {} {\emph {\bibinfo {title} {Exploring black holes:
  Introduction to general relativity}}}\ (\bibinfo  {publisher}
  {Addison-Wesley},\ \bibinfo {year} {2000})\BibitemShut {NoStop}%
\bibitem [{\citenamefont {Misner}\ \emph {et~al.}(1973)\citenamefont {Misner},
  \citenamefont {Thorne},\ and\ \citenamefont {Wheeler}}]{Misner1973}%
  \BibitemOpen
  \bibfield  {author} {\bibinfo {author} {\bibfnamefont {C~W}\ \bibnamefont
  {Misner}}, \bibinfo {author} {\bibfnamefont {K~S}\ \bibnamefont {Thorne}}, \
  and\ \bibinfo {author} {\bibfnamefont {J~A}\ \bibnamefont {Wheeler}},\
  }\href@noop {} {\emph {\bibinfo {title} {Gravitation}}}\ (\bibinfo
  {publisher} {Macmillan},\ \bibinfo {year} {1973})\BibitemShut {NoStop}%
\bibitem [{\citenamefont {Landau}\ and\ \citenamefont
  {Lifshitz}(1980)}]{Landau1980}%
  \BibitemOpen
  \bibfield  {author} {\bibinfo {author} {\bibfnamefont {L~D}\ \bibnamefont
  {Landau}}\ and\ \bibinfo {author} {\bibfnamefont {E~M}\ \bibnamefont
  {Lifshitz}},\ }\href@noop {} {\emph {\bibinfo {title} {The classical theory
  of fields}}},\ \bibinfo {edition} {4th}\ ed.,\ \bibinfo {series} {Course of
  Theoretical Physics}, Vol.~\bibinfo {volume} {2}\ (\bibinfo  {publisher}
  {Butterworth-Heinemann},\ \bibinfo {year} {1980})\BibitemShut {NoStop}%
\bibitem [{\citenamefont {Claudel}\ \emph {et~al.}(2001)\citenamefont
  {Claudel}, \citenamefont {Virbhadra},\ and\ \citenamefont
  {Ellis}}]{Claudel2001}%
  \BibitemOpen
  \bibfield  {author} {\bibinfo {author} {\bibfnamefont {C-M}\ \bibnamefont
  {Claudel}}, \bibinfo {author} {\bibfnamefont {K~S}\ \bibnamefont
  {Virbhadra}}, \ and\ \bibinfo {author} {\bibfnamefont {G~F~R}\ \bibnamefont
  {Ellis}},\ }\bibfield  {title} {\enquote {\bibinfo {title} {The geometry of
  photon surfaces},}\ }\href@noop {} {\bibfield  {journal} {\bibinfo  {journal}
  {J. Math. Phys.}\ }\textbf {\bibinfo {volume} {42}},\ \bibinfo {pages} {818}
  (\bibinfo {year} {2001})},\ \Eprint
  {http://arxiv.org/abs/arXiv:gr-qc/0005050} {arXiv:gr-qc/0005050} \BibitemShut
  {NoStop}%
\bibitem [{\citenamefont {Vieira}\ and\ \citenamefont
  {Brentan}(2016)}]{Vieira2016}%
  \BibitemOpen
  \bibfield  {author} {\bibinfo {author} {\bibfnamefont {R~S}\ \bibnamefont
  {Vieira}}\ and\ \bibinfo {author} {\bibfnamefont {H~B}\ \bibnamefont
  {Brentan}},\ }\bibfield  {title} {\enquote {\bibinfo {title} {Covariant
  theory of gravitation in the framework of special relativity},}\ }\href@noop
  {} {\  (\bibinfo {year} {2016})},\ \Eprint
  {http://arxiv.org/abs/arXiv:1608.00815} {arXiv:1608.00815} \BibitemShut
  {NoStop}%
\bibitem [{\citenamefont {Resnick}(1968)}]{Resnick1968}%
  \BibitemOpen
  \bibfield  {author} {\bibinfo {author} {\bibfnamefont {R}~\bibnamefont
  {Resnick}},\ }\href@noop {} {\emph {\bibinfo {title} {Introduction to Special
  Relativity}}}\ (\bibinfo  {publisher} {John Wiley and Sons, Inc},\ \bibinfo
  {year} {1968})\BibitemShut {NoStop}%
\bibitem [{\citenamefont {French}\ and\ \citenamefont
  {Taylor}(1968)}]{French1968}%
  \BibitemOpen
  \bibfield  {author} {\bibinfo {author} {\bibfnamefont {A~P}\ \bibnamefont
  {French}}\ and\ \bibinfo {author} {\bibfnamefont {E~F}\ \bibnamefont
  {Taylor}},\ }\href@noop {} {\emph {\bibinfo {title} {Special relativity}}}\
  (\bibinfo  {publisher} {W. W. Norton \& Company},\ \bibinfo {year}
  {1968})\BibitemShut {NoStop}%
\bibitem [{\citenamefont {Adler}(1987)}]{Adler1987}%
  \BibitemOpen
  \bibfield  {author} {\bibinfo {author} {\bibfnamefont {C~G}\ \bibnamefont
  {Adler}},\ }\bibfield  {title} {\enquote {\bibinfo {title} {Does mass really
  depend on velocity, dad?}}\ }\href@noop {} {\bibfield  {journal} {\bibinfo
  {journal} {Am. J. Phys.}\ }\textbf {\bibinfo {volume} {55}},\ \bibinfo
  {pages} {739} (\bibinfo {year} {1987})}\BibitemShut {NoStop}%
\bibitem [{\citenamefont {Okun}(1989{\natexlab{a}})}]{Okun1989}%
  \BibitemOpen
  \bibfield  {author} {\bibinfo {author} {\bibfnamefont {L~B}\ \bibnamefont
  {Okun}},\ }\bibfield  {title} {\enquote {\bibinfo {title} {The concept of
  mass},}\ }\href@noop {} {\bibfield  {journal} {\bibinfo  {journal} {Phys.
  Today}\ }\textbf {\bibinfo {volume} {42}},\ \bibinfo {pages} {31} (\bibinfo
  {year} {1989}{\natexlab{a}})}\BibitemShut {NoStop}%
\bibitem [{\citenamefont {Okun}(1989{\natexlab{b}})}]{Okun1989B}%
  \BibitemOpen
  \bibfield  {author} {\bibinfo {author} {\bibfnamefont {L~B}\ \bibnamefont
  {Okun}},\ }\bibfield  {title} {\enquote {\bibinfo {title} {The concept of
  mass (mass, energy, relativity)},}\ }\href@noop {} {\bibfield  {journal}
  {\bibinfo  {journal} {Sov. Phys. Usp.}\ }\textbf {\bibinfo {volume} {32}},\
  \bibinfo {pages} {629} (\bibinfo {year} {1989}{\natexlab{b}})}\BibitemShut
  {NoStop}%
\bibitem [{\citenamefont {Okun}(2009)}]{Okun2009}%
  \BibitemOpen
  \bibfield  {author} {\bibinfo {author} {\bibfnamefont {L~B}\ \bibnamefont
  {Okun}},\ }\href@noop {} {\emph {\bibinfo {title} {Energy and mass in
  relativity theory}}}\ (\bibinfo  {publisher} {World Scientific},\ \bibinfo
  {year} {2009})\BibitemShut {NoStop}%
\bibitem [{\citenamefont {Jammer}(2000)}]{Jammer2000}%
  \BibitemOpen
  \bibfield  {author} {\bibinfo {author} {\bibfnamefont {M}~\bibnamefont
  {Jammer}},\ }\href@noop {} {\emph {\bibinfo {title} {Concepts of mass in
  contemporary physics and philosophy}}}\ (\bibinfo  {publisher} {Princeton
  University Press},\ \bibinfo {year} {2000})\BibitemShut {NoStop}%
\end{thebibliography}%
\end{document}